\documentclass[preprint]{aastexmod}

\begin{document}
\renewcommand{\thefootnote}{\roman{footnote}}

%\title{A stream in the dwarf satellite galaxy Andromeda II as a
%  remnant of the merger between two dwarf galaxies}
\title{The remnant of a merger between two dwarf galaxies in Andromeda II}

\author{Nicola C. Amorisco\altaffilmark{1}, N. Wyn Evans\altaffilmark{2},
  Glenn van de Ven$^3$}
\affil{$^{1}$Dark Cosmology Centre, Niels Bohr Institute, University of Copenhagen, Juliane Maries Vej 30, DK-2100 Copenhagen, Denmark; amorisco@dark-cosmology.dk}
\affil{$^{2}$Institute of Astronomy, University of Cambridge, Madingley Road, Cambridge, CB3 0HA, UK}
\affil{$^{3}$Max Planck Institute for Astronomy, K\"onigstuhl 17, 69117
Heidelberg, Germany\vspace{1cm}}
\received{20 September}
\accepted{16 December 2013}
\maketitle
\textbf{Driven by gravity, massive structures like galaxies and
  clusters of galaxies are believed to grow continuously through 
  hierarchical merging and accretion of smaller systems. 
  Observational evidence of accretion events is
  provided by the coherent stellar streams crossing the outer haloes
  of massive galaxies, such as the Milky Way$^{1}$ or Andromeda$^{2}$. At
  similar mass-scales, around $10^{11}$ solar masses in stars, further observational evidence of merging
  activity is also ample$^{3,4,5}$. 
  Mergers of lower-mass galaxies are expected within the hierarchical process of galaxy formation$^6$,
  but have hitherto not been seen for galaxies with less than about $10^9$ solar masses in stars$^{7,8}$.
  Here, we report the kinematic detection of a
  stellar stream in one of the satellite galaxies of Andromeda, the
  dwarf spheroidal galaxy Andromeda II, which has a mass of only $10^7$ solar masses in stars$^9$. The properties of the stream show
  that we are observing the remnant of a merger between two dwarf
  galaxies. This had a dramatic influence on the dynamics of the
  remnant, which is now rotating around its projected major
  axis$^{10}$. %Mergers between dwarfs of such low mass are expected
 % within the hierarchical process of galaxy formation$^{7}$, but have
 % so far eluded direct observational confirmation.  With a mass of
 % only $M(<1.5{\rm kpc})=2.5^{+3.1}_{-1.1}\times 10^8 M_{\odot}$, And
 % II is the lowest mass galaxy known to display compelling evidence of
%  a merger.  
  The stellar stream in Andromeda II illustrates the
  scale-free character of the formation of galaxies, down to the
  lowest galactic mass scales.% It proves that the build-up of dwarf galaxies  
  %is akin to that of their much larger counterparts.
  }

Andromeda~II (And~II) is, in size, the second largest dwarf spheroidal galaxy known
in the Local Group, with a half light radius$^{11}$ of about 1.2 kpc
(second only to And XIX$^{12}$). With a luminosity$^{9}$ of
$L_V=9.4\times 10^6 L_{\odot}$, it currently sits$^{13}$ at a distance of 185
kpc from its host, and at an heliocentric distance of about 650
kpc. Among the satellites of M31, And~II is one of the few for
which a spectroscopic dataset of hundreds of stars is currently available. 
These observations, made using the Deep Imaging
Multi-Object spectrograph (DEIMOS) on the Keck II telescope, revealed a strong and puzzling stellar rotation, 
so far unique among the dwarf spheroidal galaxies of the Local Group$^{10}$.
The results presented here are based on a re-analysis of the latter spectroscopic 
observations, kindly provided by N. Ho and M. Geha.

We assign reliable probabilities of membership to all spectroscopic targets,
comprising more than 700 candidate red giant branch stars, 
by allowing for the presence of both foreground contaminants from the
Milky Way and interlopers from the halo population of M31. Each of the
three coexisting components (members, Milky Way halo contaminants, M31
halo contaminants) are described by a distinct spatial and kinematical
distribution, the parameters of which are measured through a maximum
likelihood technique (see Methods). 
A Bayesian approach then allows us to estimate a 
probability of membership for each available star: we count 632 high
probability members (p$>$0.85; see Extended Data Fig.~1).

We then study the kinematical properties of And~II also through a maximum
likelihood method that fully takes into account the observational
uncertainties on the line-of-sight velocity of each available giant
star in the spectroscopic sample. We find that, within measurement
errors (median value of $6.5\ {\rm kms}^{-1}$), there are no
significant deviations from the strong rotation field of And~II. 
However, despite such a smooth mean velocity field and an
otherwise flat velocity dispersion profile (see Extended Data Fig.~2), 
we identify a drop and asymmetries in the velocity dispersion field, 
especially in the circular annulus $0.9\lesssim R/{\rm kpc}\lesssim1.7$.

In order to quantify the significance of these kinematic anomalies, we
isolate a group of stars that is defined based only on its spatially
connected location.  Fig.~1a displays the giant stars identified as
high-probability members in the spectroscopic dataset, superimposed 
on Subaru Prime Focus Camera image$^{10}$ of And~II. 
We select 134 stars over an annular, stream-like region
covering an angle of 270$^\circ$ over the body of And~II (blue
points). These are compared to a control sample of 319 stars,
comprising the remainder of the spectroscopic targets at comparable
distances from the center of And~II (red points).  Although sharing a
compatible rotational field, the stream-like region is found to be
kinematically colder. Fig.~1b 
displays histograms for the line-of-sight velocity distribution of the
available giant stars in both regions. Fig.~1c shows the
normalized generalized histograms obtained after subtracting the mean
stellar rotation field. Fig.~1d shows the probability distribution
functions of the projected velocity dispersion $\sigma$. The blue
dashed curves in Figs.~1b-d refer to the stream-like region, while the red curves
are for the control sample. The probability that the velocities of the
giant stars in the two described regions have been extracted from the
same parent line-of-sight velocity distribution is negligible
($p<3\cdot 10^{-6}$). This shows that, together with the stellar
population of And~II, an additional kinematically colder component
contributes a substantial fraction of the stars in the selected
annulus.

Among the stars in the latter spatially connected stream-like region,
we next use a Bayesian approach to identify those that are
significantly better described by the properties of the control sample
with a higher velocity dispersion, $\sigma=9.3\pm0.6\ {\rm kms}^{-1}$.  These
are likely And~II stars, and we find 14 of such high probability
contaminants ($p>0.85$), which are displayed in Fig.~1a by open blue
points. If we exclude these, the remaining 120 stars (filled blue dots)
are characterized by a much colder velocity dispersion
($\sigma\lesssim$ 3~kms$^{-1}$), the probability distribution of which
is shown by the blue full curve in Fig.~1d.  We identify these stars
with a stream that extends coherently for over 5 kpc in length, with
an average thickness of 0.3 kpc.

The stellar stream contains at least 1/10 of the luminosity of
And~II, which allows us to put a safe lower bound to the total luminosity
of the progenitor $L_V\gtrsim 10^6 L_{\odot}$ (see Methods).  Furthermore, current photometric data 
do not suggest that the color spread of the red giant branch population of the stream is
dissimilar from the considerable spread of the And~II population itself
(see Extended Data Fig.~3). Together, these point to the progenitor 
of the stream being a dwarf galaxy with a total mass not too different from And~II.
A progenitor with such properties is not unexpected: recent kinematic
studies of members of the Local Group have identified analogue dwarf
galaxies with comparably low velocity dispersions around both  
the Milky Way$^{14}$ and M31$^{15}$.

The merger had a substantial influence on the dynamics and structure of the
remnant. 
The colder stream and warmer control sample do not show statistically
significant differences in the properties of their rotation, which may
seem somewhat odd within the merger context. However, it is very
likely that the orbital angular momentum between the merging dwarfs
was much larger than the intrinsic net angular momentum of the stars
in either of the dwarfs. Combined with a mass ratio expected to be not
too far from unity, the torque exerted by the merging dwarf would have
been substantial. This is most probably responsible for stars of the
stream and control sample having the same rotation strength, as well
as the puzzling orientation around the major axis -- stellar rotation
is nearly always around the minor axis, consistent with oblate
axisymmetry (except for rotation around the major axis in some giant
elliptical galaxies caused by triaxiality).
At the same time, it is likely that more stars that once belonged to the stream's 
progenitor are lurking in the spectroscopic dataset, but cannot be clearly 
disentangled from the stellar population of AndII because 
of their lower density contrast.
The detection of such a stream also provides a natural explanation for
the peculiar extended component of old stars$^{9}$ with an effectively
constant density out to a large radius of about 1.9~kpc seen in
And~II; the merger have puffed up the remnant's stellar
population.  Although a particularly close tidal encounter$^{16}$ with
M31 may also have contributed to shaping the structure of And~II, this
remains very uncertain given the unknown proper motion of And~II and
consequent degeneracy in the modelling$^{13}$.

We measure the line-of-sight velocity of the stream as well as its
projected spatial position onto the body of And~II and use this
information to constrain its approximate orbit.  As Fig.~2 shows, a
simple model of an orbit in a spherical potential is capable of
describing the available velocity and position measurements. 
Although it is not possible to infer
the detailed properties of the gravitational potential of And~II
as the orbit is found to be almost circular, we are able to constrain the
enclosed mass (stellar and dark matter) interior to the stream as follows:
$M(<1.5{\rm kpc})=2.5^{+3.1}_{-1.1}\times 10^8 M_{\odot}$.  This
implies a mass-to-light ratio of $45^{+60}_{-20}
M_{\odot}/L_{\odot,V}$, which is typical of dwarf spheroidal
galaxies$^{17,18}$.

Even with a characteristic orbital velocity for the stream, dating the
epoch of the And~II merger remains challenging.  The survival of cold
kinematic clumps is strongly dependent on the properties of the
gravitational potential in which they orbit$^{19, 20}$.
Nevertheless, the And~II stream seems to lie on a nearly
circular orbit and never passes close to the central regions of the
galaxy, which allows it to retain coherence for a very long time. 
At the same time, dynamical friction can drag two mutually orbiting dwarf galaxies closer, causing
a merger in just a fraction of the Hubble time. 
We estimate this process requires $\gtrsim 3$ Gyr, which provides us with an approximate lower limit to associate with
the merger of the And~II system. %In the end, this only provides us with an approximate lower limit of $\gtrsim 3$\,Gyr ago to associate
%with the merger of the And~II system.

Streams of disrupted and engulfed galaxies are abundant in the halo of
the Milky Way, as memorably shown in the Sloan Digital Sky Survey's
`Field of Streams'$^{1}$ and by the disrupting Sagittarius
galaxy$^{21}$.  However, the frequency and role of accretion onto
low-mass galaxies, and in particular of dwarf-dwarf mergers, remains
unclear, given the extremely limited observational evidence.  Stellar
over densities similar to shells have been discovered in the Fornax
dwarf spheroidal galaxy$^{22}$, suggestive of a late merger
origin$^{23}$. Irregular isophotes$^{24,25}$ and a kinematically cold
spot$^{19}$ indicate that Ursa Minor has suffered recent disturbance,
most probably the accretion of a lower mass stellar system in the form of a stellar cluster.
In this respect, And~II represents a compelling case of a
dwarf-dwarf merger.

Mergers between low-mass galaxies are predicted within the
hierarchical framework of galaxy formation, but they are rare at
present times. This is particularly true for dwarf satellite galaxies:
after accretion onto their host, in this case M31, the cross-section for encounters
between previously unrelated dwarf galaxies is very low$^{26,27}$,
implying that subsequent merging activity is essentially limited to
galaxies that were closely associated before infall. This makes the
discovery of a tidal stream, originating from the engulfment of one
dwarf satellite by another, particularly remarkable.  As for merger
events preceding infall, it is estimated$^{6}$ that one in two dark
matter haloes with a virial mass of $10^{10} M_{\odot}$ have
experieced a major merger (mass ratio $\gtrsim 1/3$) between redshift
4 and 1, but data to confirm these figures are exceedingly scarce.

And~II provides direct evidence for the importance of mergers even for
the smallest and least luminous of galaxies. Just as for the largest
giant ellipticals, merging and accretion were dominant processes in the formation of the dwarf galaxies we see today.

%\bigskip
%\noindent
%{\bf Methods Summary.}
%The on-line Methods Section consists of three parts. The first part details the selection of spectroscopically targeted stars as members of And II. This is followed %by a description of the method used to extract the kinematics from the member stars and subsequent kinematic detection and characterisation of the stream. The %last part outlines the luminosity estimate of the stream and inferred constraints on its progenitor.

\bigskip
\noindent
{\bf References}

\bibliographystyle{nature}
\bibliography{paper}      

\noindent
  1.  Belokurov, V., Zucker, D., Evans N. W., et al. The Field of
  Streams: Sagittarius and its Siblings. Astrophys. J. 642, L137-141  (2006).\\
  2.  Ibata, R., Irwin, M., Lewis, G., Ferguson, A., Tanvir, N. A
  Giant Stream of Metal-rich stars in the Halo of the Galaxy
  M31. Nature 412, 49-52 (2001). \\
  3.  Mart{\'{\i}}nez-Delgado, D., Pe{\~n}arrubia, J., Gabany, R.J., et
  al. The Ghost of a Dwarf Galaxy: Fossils of the Hierarchical
  Formation of a Nearby Spiral Galaxy NGC 5907, Astrophys. J. 689,
  184-193 (2008).\\
  4.  Chonis, Taylor S.; Mart{\'{\i}}nez-Delgado, David; Gabany, R.J. et al.
  A Petal of the Sunflower: Photometry of the Stellar Tidal Stream in the Halo of Messier 63 (NGC 5055),
  Astron. J. 142, 166-181 (2011).\\
  5.  Mart{\'{\i}}nez-Delgado, D., Pohlen, M., Gabany, R.~J., et
  al. Discovery of a Giant Stellar Tidal Stream around the Disk Galaxy
  NGC 4013. Astrophys. J. 692, 955-963 (2009). \\
  6.  Fakhouri, O., Ma, C.-P., Boylan-Kolchin, M. The Merger Rates
  and Mass Assembly Histories of Dark Matter Haloes in the two
  Millennium Simulations., Mon. Not. R. Astron. Soc. 406, 2267-2278 (2010).\\
  7.  Rich, R. M., et al. A tidally distorted dwarf galaxy near NGC 4449. Nature 482,
  192-194 (2012).\\
  8.  Mart{\'{\i}}nez-Delgado, D., et al. Dwarfs gobbling dwarfs: a stellar tidal stream around
  NGC 4449 and hierarchical galaxy formation on small scales. Astrophys. J. 748,
  L24 (2012).\\
  9.  McConnachie, A. W., Arimoto, N., Irwin, M. J. Deconstructing
  Galaxies: a Suprime-Cam survey of Andromeda II. Mon. Not. R. Astron. Soc. 379, 379-392 (2007).\\
  10.  Ho, N., Geha, M., Munoz, R. R., et al. Stellar Kinematics of the
  Andromeda II Dwarf Spheroidal Galaxy. Astrophys. J. 758, 124-136 (2012). \\
  11.  McConnachie, A. W., Irwin, M. J. Structural Properties of the M31
  Dwarf Spheroidal Galaxies, Mon. Not. R. Astron. Soc. 365,
  1263-1276 (2006).\\
  12.  McConnachie, A.W.; Huxor, A.; Martin, N.F. et al.
  A Trio of New Local Group Galaxies with Extreme Properties
  Astrophys. J. 688, 1009-1020, (2008).\\
  13.  Watkins, L. L., Evans, N. W., van der Ven, G. A Census of Orbital
  Properties of the M31 Satellites. Mon. Not. R. Astron. Soc. 430, 971-985 (2013).\\
  14.  Koposov, S. E., Gilmore, G.; Walker, M. G., et al. Accurate Stellar Kinematics at Faint Magnitudes: 
  Application to the Boštes I 
  Dwarf Spheroidal Galaxy. Astrophys. J. 736, 146 (2011).\\
  15.  Collins, M. L. M., Chapman, S. C., Rich, R. M., et al. A Kinematic Study of the Andromeda Dwarf Spheroidal System. 
  Astrophys. J. 768, 172 (2013).\\
  16.  Shaya, E. J., Tully, R. B. The formation of Local Group planes of galaxies. Mon. Not.
   R. Astron. Soc. 436, 2096-2119 (2013). \\
  17.  Walker, M. Dark Matter in the Galactic Dwarf Spheroidal Satellites in
  `Planets, Stars and Stellar Systems' Vol. 5, Oswalt, T.D., Gilmore, G., Springer (2013)\\
  18.  Amorisco, N. C., Evans, N. W. Phase-space models of the dwarf spheroidals. Mon. Not. R. Astron. Soc. 411, 2118-2136 (2011).\\
  19.  Kleyna, J. Wilkinson, M. I., Gilmore, G., Evans, N. W. A
  Dynamical Fossil in the Ursa Minor Dwarf Spheroidal. Astrophys. J.
  588, L21-24 (2003).\\
  20.  S{\' a}nchez-Salcedo, F. J., Lora, V., The survival of dynamical fossils in dwarf spheroidal 
   galaxies in conventional and modified dynamics.
  Mon. Not. R. Astron. Soc. 407, 1135-1147 (2010).\\
  21.  Belokurov, V. Koposov, S., Evans, N. W., et al. Precession of the Sagittarius stream. Mon. Not. R. Astron. Soc.
  437, 116-131 (2014). \\  
  22.  Coleman, M., Da Costa, G. S., Bland-Hawthorn, J., et al. Shell
  Structure in the Fornax Dwarf Spheroidal. Astron. J. 127, 832-839 (2004). \\
  23.  Amorisco, N. C., Evans, N. W., A Troublesome Past: Chemodyanmics
  of the Fornax Dwarf Spheroidal, Astrophys. J. 756, L2 (2012). \\
  24.  Irwin, M., Hatzidimitriou, D., Structural parameters for the Galactic dwarf spheroidals. 
  Mon. Not. R. Astron. Soc. 277, 1354-1378 (1995).\\
  25.  Palma, C., Majewski, S. R., Siegel, M. H., et al., Exploring Halo Substructure with Giant Stars. IV. 
  The Extended Structure of the Ursa Minor Dwarf    
  Spheroidal Galaxy. Astron. J. 125, 1352-1372 (2003).\\
  26.  Tremaine, S., `Galaxy Mergers' in `Structure and Evolution of
   Normal Galaxies' (eds S.M. Fall, D. Lynden-Bell) pgg. 67-84, Cambridge University
  Press (1981).\\
  27.  De Rijcke, S.; Dejonghe, H.; Zeilinger, W. W.; Hau, G. K. T. 
  Dwarf elliptical galaxies with kinematically decoupled cores, 
  Astron. Astrophys. 426, 53-63 (2004).\\
 
\bigskip
\noindent
{\bf Acknowledgments}. The authors are 
glad to thank Mike Irwin for discussions on the photometric properties
of And~II. The Dark Cosmology Centre (DARK) is founded by the Danish
National Research Foundation (DNRF). This work was partially supported
by Sonderforschungsbereich SFB 881 ``The Milky Way System''
(subproject A7) of the German Research Foundation (DFG).

\bigskip
\noindent
{\bf Author contributions}. NCA performed the candidate selection
using methods originally developed with NWE, and the subsequent
kinematic extraction together with GvdV. NCA, NWE and GvdV suggested
and elaborated the stream model to explain the data. The paper was
written by NCA, with contributions from the other two authors.

\bigskip
\noindent
{\bf Corresponding author}. Nicola C. Amorisco amorisco@dark-cosmology.dk

\begin{figure*}
\figurenum{{\bf 1}}
\centering
\includegraphics[width=.7\textwidth]{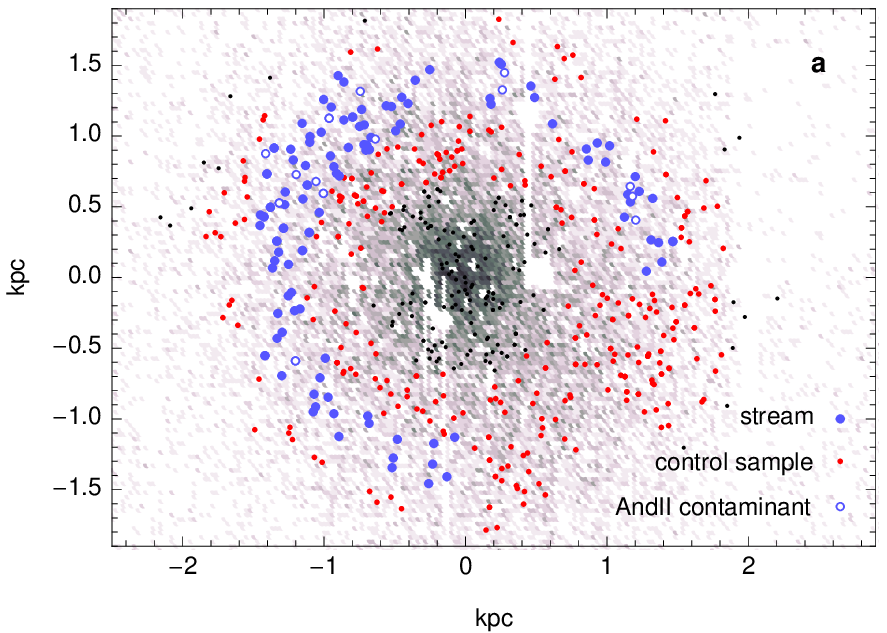}\\
\includegraphics[width=.3\textwidth]{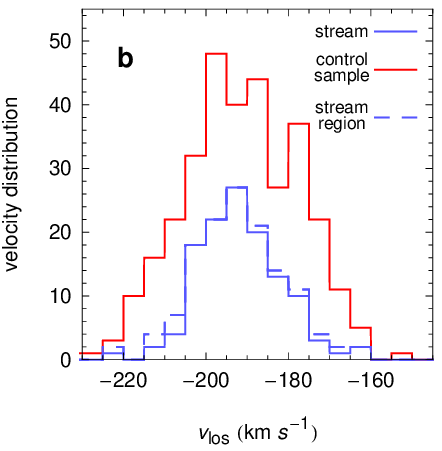}
\includegraphics[width=.31\textwidth]{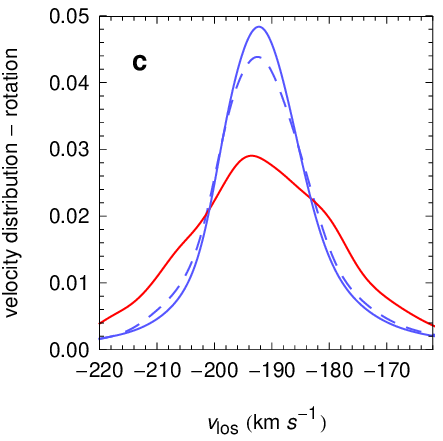}
\includegraphics[width=.305\textwidth]{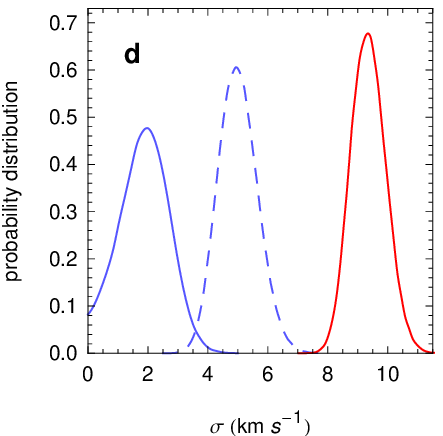}
\caption{
{\bf Figure 1 $|$ Kinematic detection of a stream in And~II.}  
The targets in the circular annulus $0.7<R/{\rm kpc}<1.9$ are divided 
into a kinematically cold and a warm component, panel {\bf a}. The 134 stars in the connected stream-like
region (blue points) yield the blue-dashed line-of-sight velocity
distributions in panels {\bf b} and {\bf c} (respectively before and 
after normalisation, subtraction of the mean stellar rotation field and convolution
with the individual measurement uncertainties), and the blue-dashed probability
distribution for the projected velocity dispersion $\sigma$ in panel
{\bf d}. Red distributions are associated with the kinematically warmer control
sample, comprising the remaining 319 spectroscopic targets (red
points). 
The 14 blue open points isolate stars which are more likely to belong 
to the main body of And~II rather than to the stream itself. When they 
are subtracted from the sample of stream stars, the internal velocity 
dispersion of the stream is reduced further, leading to the blue full distributions.}
\end{figure*}  
  
\bigskip
\noindent
\begin{figure*}
\centering
\includegraphics[width=.7\textwidth]{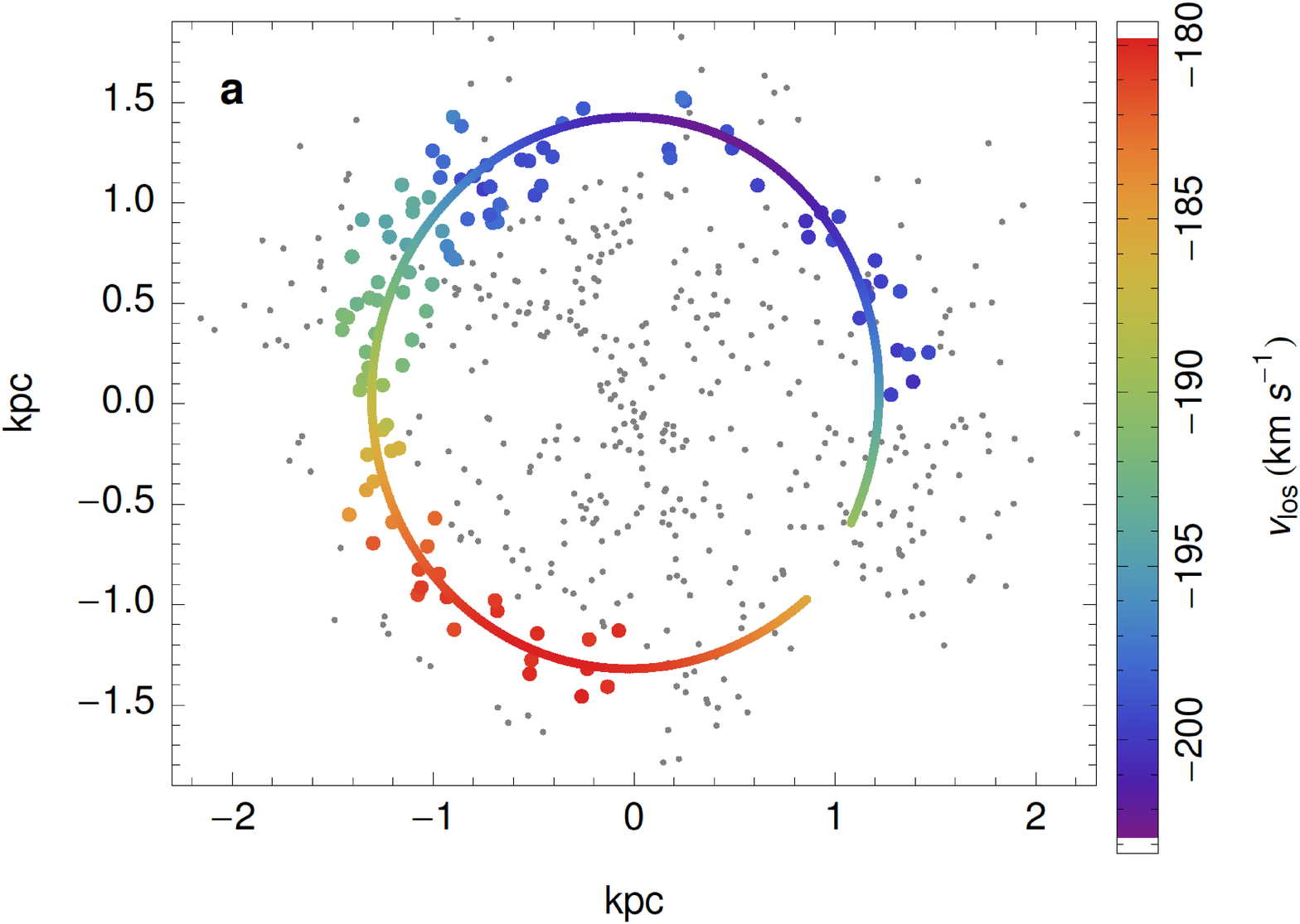}\\
\vspace{.2in}
\includegraphics[width=.35\textwidth]{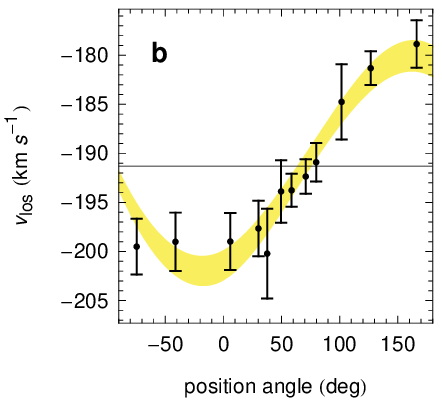}
\includegraphics[width=.326\textwidth]{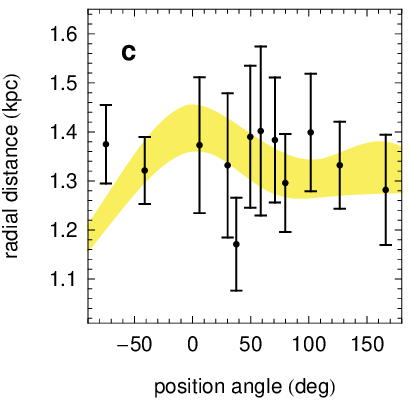}
\caption{
{\bf Figure 2 $|$ The stream reproduced by an orbit in a spherical
  potential.}  The line-of-sight velocity (panel {\bf b}) and distance
from the centre of Andromeda II (panel {\bf c}) of the stars belonging
to the stream are reproduced by a simple model of an orbit in a
spherical potential. Panel {\bf a} displays the spectroscopic targets
belonging to the stream, color-coded using a smoothed velocity field
for comparison with the best-fitting orbit. Panels {\bf b} and {\bf c}
show a direct comparison between the observables (each datapoint has
been obtained by a subsets of 10 stream stars, error bars are s.d.) and
the corresponding 68$\%$ confidence region obtained from the model.}
\end{figure*}

\clearpage
\noindent
{\bf Methods}

\noindent
{\bf I -- Membership Selection.}
\smallskip\\
\noindent
In order to address the
kinematical properties of And II, we need to separate the spectroscopic
targets that belong to the red giant branch population of the dwarf
galaxy from any contaminants. Contamination has two different origins:
dwarf stars in the foreground belonging to the Milky Way and
interlopers from the stellar halo of Andromeda.  To reliably
disentangle these three distinct components, we model the spectroscopic
dataset as a superposition of multiple independent stellar
populations, within the framework of a maximum likelihood
technique$^{23, 28}$:
\begin{equation}
L({\vec \Theta})=\prod_j \sum_i f_i\ P^{\rm sp}({\vec X}_j; {\vec \Theta}^{\rm sp}_i)\ P^{\rm kin}(v_j; {\vec \Theta}^{\rm kin}_i)\ .
\label{eq:popsmodel}
\end{equation}  
The index $j$ runs over the spectroscopic targets, while $i$ indicates
the three populations of the model, each containing a fraction of
stars $f_i$, with the constraint $\sum_i f_i=1$. The probability
function $P^{\rm sp}_i$, parametrized by the set of parameters ${\vec \Theta}^{\rm sp}_i$, describes the spatial distribution of the
members of the component $i$ on the plane of the sky.  Given the
limited angular size of And II, we can adopt a constant surface
density distribution for both populations of contaminants, while a
Plummer density profile with elliptical isophotes is used to describe
the population of And II itself.  Analogously, the three components
have different probability distributions for their kinematics, which
are described by the function $P^{\rm kin}$ and the parameters ${\vec \Theta}^{\rm kin}_i$. Given the significant
separation of the contaminants from And II in terms of systematic
velocity$^{10}$, we can describe both the Milky Way and the Andromeda
halo population with a simple Gaussian line-of-sight velocity
distribution.  The observational uncertainty on the measurement of the
line-of-sight velocity of each single spectroscopic target,
$\delta_j$, is fully included in the analysis by formal convolution;
hence, for the contaminants, we have:
\begin{equation}
P_{\rm cont}^{\rm kin}(v_j; v^{\rm sys}_i, \sigma_i)={1\over{\sqrt{2\pi(\sigma_i^2+\delta_j^2)}}}{\exp\left[-{1\over 2} {{(v_j-v^{\rm sys}_i)^2}\over{\sigma_i^2+\delta_j^2}}\right]} \ ,
\label{eq:losvdcont}
\end{equation}
where Milky Way and Andromeda contaminants have their own systematic
velocity $v_i^{\rm sys}$ and intrinsic velocity dispersion $\sigma_i$.  
The velocity distribution of And II is also normally distributed, but has a
rotating mean velocity field:
\begin{equation}
P^{\rm kin}_{\rm And II}(v_j; v^{\rm sys}, \vec{\Omega}, \sigma)={1\over{\sqrt{2\pi(\sigma^2+\delta_j^2)}}}{\exp\left[-{1\over 2} {{(v_j-v^{\rm sys}-\vec{\Omega}\cdot \vec{X}_j)^2}\over{\sigma^2+\delta_j^2}}\right]}\ .
\label{eq:losvdAndII}
\end{equation}

The parameter space is explored by means of a suite of Monte Carlo
chains, constructed using the Metropolis-Hastings
algorithm$^{29}$. Once the best-fitting model is identified, for
example as described by the set of parameters $\vec{\Theta}^{\rm bf}$,
a Bayesian approach allows us to estimate, for each spectroscopic
target $j$, the probability of that individual star belonging to the
population $i$:
\begin{equation}
p_j^i = {{f_i\ P^{\rm sp}({\vec X}_j; {\vec \Theta}^{\rm bf,sp}_i)\ P^{\rm kin}(v_j; {\vec \Theta}^{\rm bf,kin}_i)}
\over{\sum_k f_k\ P^{\rm sp}({\vec X}_j; {\vec \Theta}^{\rm bf,sp}_k)\ P^{\rm kin}(v_j; {\vec \Theta}^{\rm bf,kin}_k)}}\ .
\label{eq:memprop}
\end{equation}
Extended Data Fig.~1 shows the distribution of all spectroscopic targets in the plane $(v_{los}, R)$, where $R$ is the
projected radial distance from the centre of And II. Color-coding of
each target is set by its probability of membership to And II: the
combination of spatial position and kinematical information is capable
of efficiently selecting the members of And II. As a further check, we
also exclude a few targets that appear as fortuitous `kinematic
members', but whose membership to And II is questioned$^{10}$ either by their
colour ($V-I>2.5$) or by the strength of the Na~I line (EW$_{\rm Na~I}>4$),
which indicate that these may in fact be dwarf foreground stars from
the Milky Way halo. As a result, we identify 632 giant stars belonging to And II with high probability ($p>0.85$).

\bigskip
\noindent
{\bf II -- Kinematic extraction.}
\smallskip\\
\noindent
Despite significant observational effort and use of the DEIMOS spectrograph on
Keck, the heliocentric distance of And II, $D\approx650$ kpc, is so
great that uncertainties on the measurement of the line-of-sight
velocity of each target are significantly higher than figures achieved
for the closer Milky Way dwarf satellites ($\delta_v\approx 2$
kms$^{-1}$).  In particular, with a median of about
$\delta_v\approx6.5$ kms$^{-1}$, measurement errors are comparable in
magnitude to the kinematical spread due to the intrinsic velocity
dispersion of the dwarf, which implies that appropriate treatment of
these uncertainties is crucial.

For this reason, we extract the kinematic properties of And II by
fully taking into account all measurement errors, individually for
each spectroscopic target.
First, we adopt the line-of-sight velocity distribution of Equation~\ref{eq:losvdAndII} to
infer the intrinsic velocity dispersion $\sigma$ in circular annuli as displayed in Extended Data Fig.~2. 
The resulting $\sigma$ profile is similarly flat as typically observed in dwarf spheroidal galaxies,
except for a significant dip around $R \simeq 1.3$\,kpc, independent of the precise choice of the circular annuli.

Next, we applied the same approach to stars in the circular annulus $0.9\lesssim R/{\rm kpc}\lesssim1.7$, 
but now allowing for two spatially connected components with a different intrinsic velocity dispersion $\sigma$.
This results in the identification of the stream-like region indicated by blue circles in panel \textbf{a} of Fig.~1 in the Letter.
The convolution of the intrinsic line-of-sight velocity distribution with the assumed normally distributed error function of each
target is illustrated in panel \textbf{c} of Fig.~1, which displays normalised generalised histograms for the
velocity distributions of both stream and control sample.
Only by explicitly including each measurement error individually it is then possible to infer the 
probability distributions of the intrinsic dispersion for stream and control sample, 
as shown in panel \textbf{d} of Fig.~1. Well over $10^5$ draws in a suite of Monte Carlo chains 
have been used in order to accurately
sample the tails of these $\sigma$ distributions. This technique also ensures that these $\sigma$ 
distributions are properly marginalised against
uncertainties of all other parameters of the model.

\bigskip
\noindent
{\bf III -- Stream luminosity and progenitor.}
\smallskip\\
\noindent
Our kinematical analysis identifies $N_{\rm str}=120$ red giant stars as
high probability members of the kinematically cold stream, in a pool
of $N_{\rm mem}=632$ members of AndII. Although indicative of the
significant luminosity of this structure, it is not entirely correct
to simply use the ratio $N_{\rm str}/N_{\rm mem}$ to derive an
estimate of the luminosity of the stream itself. This is because the
stream covers a limited area over the body of And II and the
spectroscopic coverage is neither uniform over the body of the dwarf
nor proportional to its surface brightness profile.

We can instead obtain more accurate insight into the luminosity of the
stream by restricting ourselves to the projected region it
covers. This has been identified in Fig.~1a as an approximately annular
region, centred around $R\approx1.35$ kpc and covering an angle of
$\approx270^{\circ}$. By using the detailed surface density
profile$^{9}$ of And II, obtained by using deep Subaru Suprime-Cam
data, we calculate that such projected area contains a fraction of
$\approx15\%$ of the total luminosity of the dwarf. In turn, we only
find 14 high probability contaminants in this region, i.e. stars that
are significantly more likely to belong to the And II population
rather than to the kinematically cold stream. This implies a
luminosity estimate for the stream of
\begin{equation}
L_{V, {\rm str}} \approx  0.15 \times {120\over134} \times L_{V, {\rm And II}} 
\approx 0.13 \times L_{V, {\rm And II}} \approx 1.3\times 10^6 \, L_{\odot}
\end{equation}

It is worth mentioning that, based on kinematical data alone, we are
inevitably only capable of identifying those parts of the stream which
have a sufficiently high density contrast against the And II
population, and whose kinematics in the stream's progenitor were cold
enough to stand out against the average current properties of And
II. In particular, as a result of the merger, it is plausible that any
stellar components that were originally more diffuse in the stream's
progenitor are now dispersed in And II, eluding a purely kinematical detection. 
This implies that the above 13\% should be regarded as a lower limit for the luminosity of the
progenitor itself.

Whereas the resulting luminosity already makes a stellar cluster origin unlikely, 
this can be excluded further by considering the properties of the distribution of
the stream's members in the Colour Magnitude Diagram (CMD). Extended Data Fig.~3 follows the same 
stream versus control sample labeling as in panel \textbf{a} of Fig.~1 in the Letter to 
allow for the comparison of the properties of the stream's member stars with the And II population. 
We find that this data suggest little difference in the distribution of such two populations in the CMD, 
highlighting that the stream also has a significant color spread.

\noindent
{\bf Methods References.}

\noindent
  28. Walker, M.~G., \& Pe{\~n}arrubia, J. A Method for Measuring (Slopes of) the Mass 
  Profiles of Dwarf Spheroidal Galaxies. Astrophys.  J. 742, 20 (2011).\\
  29. Metropolis, N., Rosenbluth, A.W., Rosenbluth, M.N., Teller, A.H., Teller, E. Equation of state 
  calculations by fast computing machines. J. Chem. Phys., 21, 1087-1092 (1953).\\

\noindent
\begin{figure}
\centering
\includegraphics[width=.7\textwidth]{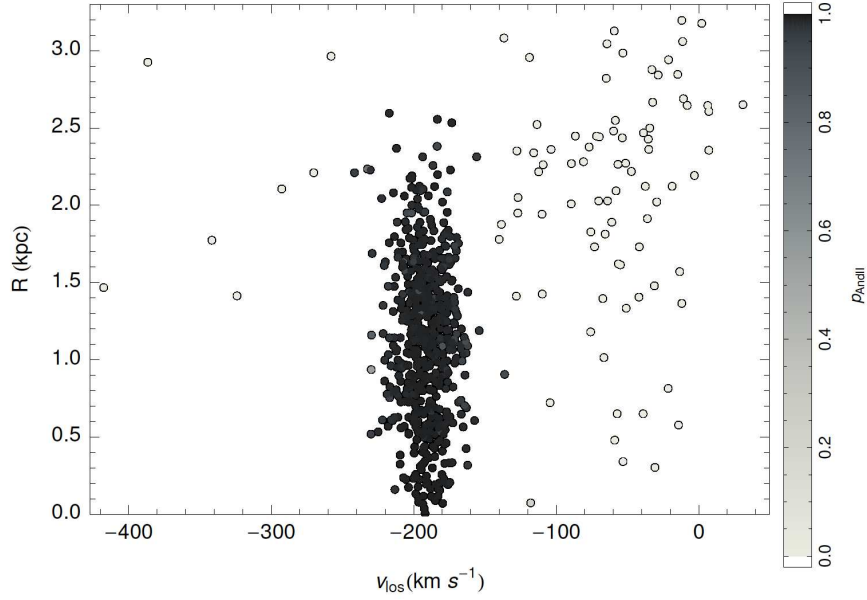}
\caption{
{\bf Extended Data Figure 1 $|$ Membership Selection.}
The spectroscopic
dataset in the plane $(v_{los}, R)$, colour-coded according to the
probability of each target belonging to the stellar population of And
II. Non-member targets with velocities higher than the systematic
velocity of And II ($v_{\rm sys}=-191.4\pm0.4$) are foreground
contaminants from the Milky Way, while non-member targets at lower
negative velocities are interlopers from the Andromeda halo.}
\end{figure} 

\noindent
\begin{figure}
\centering
\includegraphics[width=.55\textwidth]{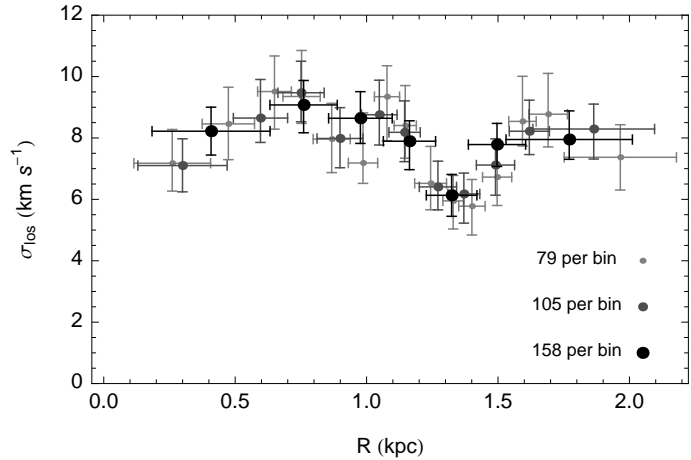}
\caption{
{\bf Extended Data Figure 2 $|$ Velocity dispersion profile.}
 And~II displays an approximately flat velocity dispersion
profile, except for a significant dip near the average projected radius of the stellar stream. 
Points of different sizes and color-dephts refer to different circular annuli sizes (as in the legend), while error bars display 68$\%$ confidence levels.}
\end{figure}

\noindent
\begin{figure}
\centering
\includegraphics[width=.55\textwidth]{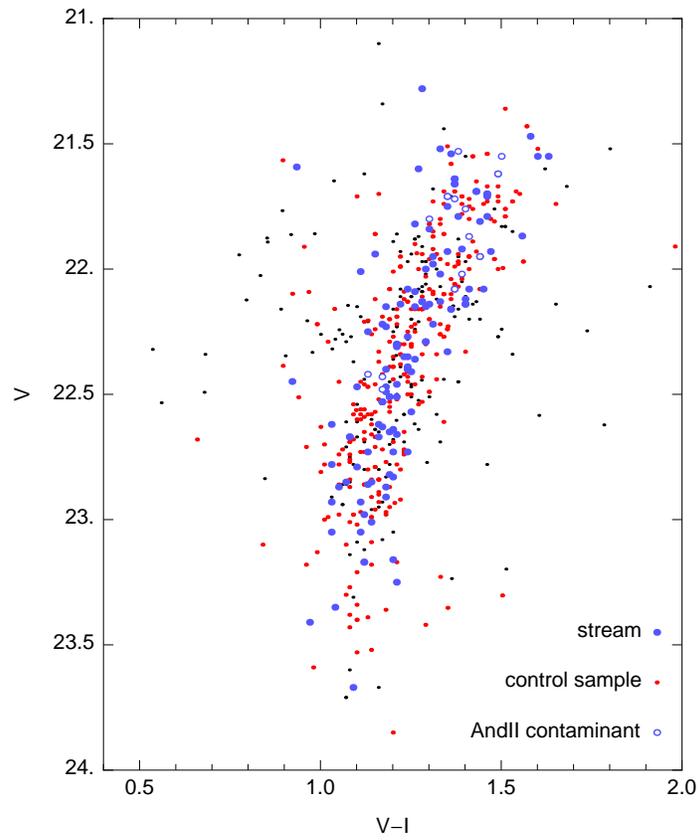}
\caption{
{\bf Extended Data Figure 3 $|$ Color Magnitude Diagram.}
The distribution of the stars belonging to the stream (blue points) and
stars in the control sample (red points) in V-band magnitude versus V-I color.}
\end{figure}

\end{document}